\documentclass[]{article}
\usepackage{cite}
\usepackage{amsmath,amssymb,amstext,times}
\usepackage{color}
\usepackage{float}
\usepackage{pdfpages}
\usepackage{tcolorbox}
\usepackage{tabularx}
\usepackage{array}
\usepackage{colortbl}
\usepackage{arydshln}
\usepackage{array,multirow}
\usepackage{setspace} 
\DeclareMathOperator*{\argmax}{argmax}

\title{Edge-based LBP description of surfaces with colorimetric patterns}
\author{E. Moscoso Thompson and S. Biasotti}

\begin{document}
	
\maketitle
\begin{abstract}
In this paper we target the problem of the retrieval of colour patterns over surfaces. We generalize to surface tessellations the well known Local Binary Pattern (LBP) descriptor for images. The key concept of the LBP is to code the variability of the colour values around each pixel. In the case of a surface tessellation we adopt rings around vertices that are obtained with a sphere-mesh intersection driven by the edges of the mesh; for this reason, we name our method edgeLBP. Experimental results are provided to show how this description performs well for pattern retrieval, also when patterns come from degraded and corrupted archaeological fragments.
\end{abstract}

\section{Introduction}
Thanks to advances in the modeling techniques and to the availability of cheaper yet effective 3D acquisition devices, we see a remarkable increase of the amount of 3D data available. Many sensors are able to acquire not only the 3D shape but also its \emph{texture}; this is the case, for instance, of the Microsoft Kinect device. 
The creation of an increasing number of 3D models has opened new opportunities to study the past, by giving access to plenty of representations of artifacts close to their original form. At the same time, Cultural Heritage owns a growing mass of non-interpreted 3D data, which call for innovative solutions for the analysis of data. In this context, local descriptors, feature recognition and similarity measures become indexes to the informative content of 3D models, and are essential to categorize objects and to recognize a style, e.g. to attribute objects to a given society or to a given author.
A typical problem the archaeologists face when dealing with collections of fragments is to determine their compatibility. 
Compatibility is generally determined by multiple factors: geometric correspondence, same material and, possibly, if there are not evidently matching fragments, continuity consideration on the fragment skin (colour, texture) \cite{gravitate_proc}. 

Within the large scenario of Cultural Heritage, we focus on the analysis and description of color patterns. The idea is to recognize the same decoration, for instance a repeated lotus leaf, independently of the support (e. g., the surface bending) on which it is depicted. Therefore, this work will contribute to the definition of a compatibility measure among artifacts based on skin decorations. To approach this problem, we consider a novel extension of the Local Binary Pattern description to surface tessellations based on the evolution of the color over concentric circles around a vertex. To determine these circles we adopt a sphere - edge intersection strategy and for this reason we name our approach edgeLBP. As  application of the edgeLBP description, we propose the retrieval and classification of color patterns over surfaces.

The remainder of the paper is organized as follows.
Section \ref{sec:star} briefly reviews the literature on the retrieval of textured images and surfaces. Section \ref{sec:method}  introduces  the  elements  of  our  method,  i.e.  the edgeLBP operator and how we store it in a descriptor. Section \ref{sec:results} presents and analyses the retrieval and classification performances of the method over two datasets, while conclusive remarks end the paper, Section \ref{sec:conclusions}.
\section{State of art}
\label{sec:star}
A typical strategy to detect textures on images is to consider local patches that describe the behavior of the texture around a group of pixels. Examples of these descriptions are the Local Binary Patterns (LBP) \cite{ojala,ojala02}, the Scale Invariant Feature Transform (SIFT) \cite{Lowe2004} and the Histogram of Oriented Gradients (HOG) \cite{DaTr05}.
The generalization of these descriptions to (even textured) surfaces has been explored in several works, such as  the PANORAMA views of the 3D objects \cite{Papadakis2010}, the meshHOG \cite{meshHOG} and the meshLBP \cite{WerghiTBB16,WerghiTBB15}.
%
%
In general, the methods for matching textured 3D shapes adopt a combination of geometric and colorimetric descriptors. Possible choices of the colorimetric descriptors are: feature-vectors, where  the color is treated as a general property of the shape, \cite{Suzuki01}, or its subparts in \cite{garro16}; local or global views of the objects \cite{WuCLFP08,Pasqualotto2013}; point-to-point correspondences among sets of feature points (e.g., the CSHOT descriptor \cite{TombariSS11}); the evolution of the sub-level sets according to the persistent homology settings \cite{PHOG}. 
These methods mainly address the shape matching problem without focusing on the surface details and local colorimetric variations. On the contrary, when looking for patterns, locality and scale are the two key aspects.
A detailed evaluation and comparison of methods for 3D texture retrieval and comparison can be found in \cite{Biasotti2016} and several SHREC contests \cite{Cerri13,Cerri14,Giachetti15}. However, all these contests focused on the joint comparison of geometry and texture, without considering the comparison of the purely colorimetric information that characterizes the surface decorations. 

At the best of our knowledge, the Mesh Local Binary Pattern (meshLBP) approach \cite{WerghiTBB16,WerghiTBB15,WerghiBB15} is the unique approach that explicitly addresses pattern analysis over surfaces. The meshLBP extends the LBP \cite{ojala} to triangle meshes. The main idea behind the meshLBP is that triangles play the role of pixels and the 8-neighbor connectivity in an image is ideally substituted by a 6-neighbor connectivity around triangles. Rings on the mesh are computed using a uniform, region growing, triangle-based expansion. 
From the practical point of view, the meshLBP encodes a pattern efficiently, providing a compact representation of it.
\section{The edgeLBP}
\label{sec:method}
We extend the LBP to surfaces using rings defined on the basis of a sphere-mesh intersection. 
In Section \ref{ss:LBP_imagesc} we briefly sum up the definition of the LBP definition. Our extension to surface tessellations is described in Section \ref{ss:edgeLBP_def}, while Section \ref{ss:descriptor} details the edgeLBP descriptor and the distance adopted to compare two descriptors.
\subsection{Local Binary Pattern for gray-scale images}
\label{ss:LBP_imagesc}
\begin{figure}
\begin{center}\begin{tabular}{ccc}
\includegraphics[width=3.5cm]{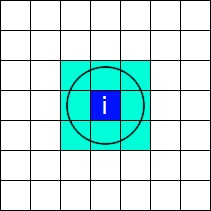}&
\includegraphics[width=3.5cm]{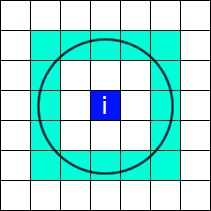}&
\includegraphics[width=3.5cm]{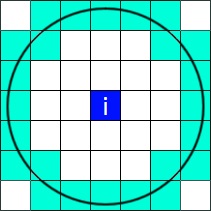}\\
(a)&(b)&(c)
\end{tabular}
\caption{In (a) the $ring$ of the pixel $i$ is shown; while (b) and (c) show two examples of concentric $ring$s.}
\label{fig:LBP_ring}
\end{center}
\end{figure}

The \emph{Local binary pattern (LBP)} and its variants prove to be a good solution for the classification of patterns in images \cite{LIU2017}. Given a gray-scale image $I$, the LBP describes the pattern in $I$ coding the local variation of the gray-scale values (encoded with a function $h:I\rightarrow [0,255]$) around each pixel of $I$. More extensively, for each pixel $i\in I$, a \emph{ring} of pixels around $i$ (called $ring_i$) is considered (see Figure \ref{fig:LBP_ring}) and a 8-digit binary array $str_i$ defined as follow:
$$str_i(j)=\Bigl\{	\begin{array}{ll}	1 & if \quad h(i)<h(i_j)\\
										0 & otherwise			\\
    			  	\end{array}$$
where $i_j$ is the $j-th$ pixel of the ring around $i$, sorted clockwise and starting from the top-left pixel.
The LBP operator of a pixel $i$ is defined by: $$LBP(i)=\sum_j str_i(j)\alpha(j),$$ where $\alpha$ is a weight function. Throughout this paper we consider $\alpha_1(j)=1, \forall j$. Notice that in this case, the $LBP(i)$ value is independent of the ordering of $ring_i$. Finally, the LBP descriptor of the pattern in $I$ is defined as the histogram of the values $LBP(i)$. 

The LBP operator was extended to multiple rings around each pixel in $I$, see Figure \ref{fig:LBP_ring}(b-c). The descriptor of the LBP multi-ring is the concatenation of the histograms of the LBP values of each single ring, e.g., an array or a matrix.
\subsection{Definition and implementation of the edgeLBP operator}
\label{ss:edgeLBP_def}
We extend the multi-ring LBP operator to deal with surface tessellations through a sphere-mesh intersection technique, called the \emph{edge Local Binary Pattern (edgeLBP)}. By a surface tessellation, we mean a polygonal mesh $T=(V,E,F)$, which is a collection of vertices $V$, edges $E$ and faces $F$ defining the surface of an object. In our settings, we assume that the faces of the tessellation are convex polygons; examples of admissible surface representations are triangle and quad meshes, \cite{BLPPSTZ13a}.

We assume that the surface property can be stored as a scalar function $h$ defined on the vertices of the tessellations, formally, $h:V\rightarrow\mathbb{R}$. In our settings, we consider two choices for the function $h$: (i) the \emph{L-channel} from the CIELab color space \cite{CIELAB_1,CIELAB_2}; (ii) the gray-scale value defined as $0.21R+0.72G+0.07B$ ($R$, $G$ and $B$ are the channels of the RGB color space).

The concept of ring is crucial for the LBP operator: while a pixel grid has the same connectivity everywhere, surface tessellations can be widely \emph{irregular}, thus the $ring$ definition over them is not obvious. By irregular we mean that the vertices can be non uniformly distributed over the surface and the faces of the tessellation may have different area, shape and number of edges.
\begin{figure}
\begin{center}\begin{tabular}{cc}
\includegraphics[width=5.5cm]{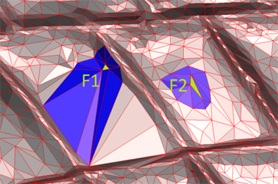}&
\includegraphics[width=5.5cm]{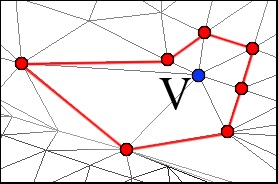}\\
(a)&(b)
\end{tabular}
\caption{(a): in blue, two rings defined on the basis of triangles; (b): the ring around the vertex $v$ is defined by mesh vertices (red dots).}
\label{fig:bad_rings}
\end{center}
\end{figure}
%
Figure \ref{fig:bad_rings} depicts two possible $ring$ definitions exclusively made of mesh elements (triangles in Figure \ref{fig:bad_rings}(a) and vertices in Figure \ref{fig:bad_rings}(b), resp.): in both cases, the irregularity of the mesh elements strongly influences these of rings.

We define the $ring$ of a vertex $v \in V$ as the intersection of the surface tessellation with a sphere of radius $R$ centered in $v$. 
\begin{figure}
\centering
\begin{tabular}{cc}
\includegraphics[width=5.5cm]{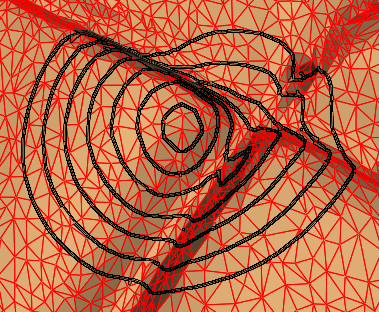}&
\includegraphics[width=5.5cm]{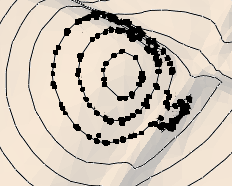}\\
(a)&(b)
\end{tabular}
\caption{(a): in black, multiple closed curves defined by the set of points $P_i \in \mathcal{R}$; (b): the black dots correspond to the elements $p_i$ of the three central curves in (a).}
\label{fig:ring_construction}
\end{figure}
Such an intersection is represented by the set of points $\mathcal{R}=\{p_1, p_2, \ldots, p_k\}$ that approximate the intersection between the sphere and the surface. 
Figure \ref{fig:ring_construction} shows a number of concentric rings over a triangle mesh. To determine a $ring$ around a vertex $v$, we follow a mesh expansion approach driven by the Euclidean distance from the vertex $v$, as summarized in the following steps:
\begin{enumerate}
\item All the edges that are incident in $v$ are added to a list $L$.
\item Starting from an edge $e=(v,v_1) \in L$, the intersection between $e$ and the sphere centered in $v$ with radius $R$ is evaluated. If there actually is an intersection, it is stored as a new point $p_i$, otherwise, if $e$ completely falls inside the sphere, we add to $L$ all the edges that are incident to $v_1$. The edge $e$ is removed from $L$ and labeled as visited.The value $h(p_i)$ on $p_i$ is given by the linear interpolation of the values that $h$ assumes in $v$ and $v_1$. 
\item The step $2$ is repeated $\forall e \in L$, until the list is empty.
\end{enumerate}

To achieve a multi-ring representation, for any vertex $v \in V$ we consider $N_r$ rings, $\{ring_1^v, \ldots, ring_{N_r}^v\}$. Let $S^v_{l}$ be the surface portion of $T$ that contains $v$ and has the $ring^v_k$ as its boundary, $l=1\ldots N_r-1$, then the relation $S^v_{l} \subset S^v_{l+1}$ holds for each $l$.
When extending the edgeLBP evaluation to multiple rings, the algorithm takes advantage of the nested nature of the rings and extracts $S^v_l$ with respect to increasing values of the radius $R$. 


In general, the sphere-surface intersection can produce multiple, closed curves that bound either a multiple connected or a dis-connected portion of the surface, as detailed in \cite{blowing_bubbles}.
Using a region growing approach, we dynamically consider only the $S^v_l$ components.
Therefore, $S^v_l$ is always a connected region that contains $v$; however, it can become multiply connected. If all the $N_r$ components of $S^v_l$ are simply connected and all the $N_r$ rings do not intersect the surface boundary (if any), the 
$v$ is considered an \emph{admissible} vertex for the edgeLBP, otherwise it is \emph{non-admissible}.

\subsubsection{Ring re-ordering and sampling}
\begin{figure}
\centering
\includegraphics[width=12cm]{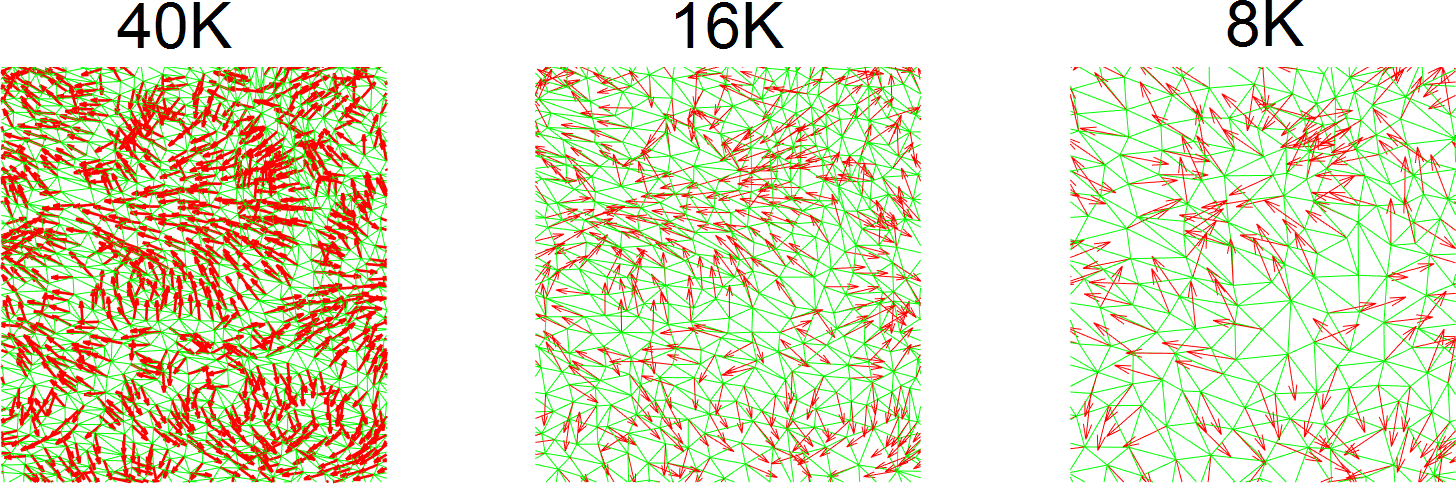}
\caption{Arrows represent the starting point of the rings in meshes representing the same surface but sampled with a different number of vertices (40K, 16K and 8K vertices, resp.).}
\label{fig:ring_stab}
\end{figure}

Each $ring$ is represented as the piecewise, linear curve $C$ determined by the segments $(p_i,p_{i+1})$, $p_i \in \mathcal{R}$.
%
Then, the curve $C$ is oriented \emph{counter-clockwise} with respect to the vector in $v$ normal to $T$. We select 
As the starting point for ordering $C$, we select the point $\tilde{p}$ such that: 
$$\tilde{p}=\argmax\limits_{p_i\in\mathcal{R}} h(p_i).$$
In case of symmetries around a vertex, multiple choices of the starting point are possible: we select the candidate point that is the farthest from the other elements of $\mathcal{R}$.
The stability of the starting point of a $ring$ is confirmed in numerous experiments we performed on meshes of different resolution, where by mesh resolution we mean the number of vertices of the mesh.
Figure \ref{fig:ring_stab} shows the vector field generated by the difference between $\tilde{p}$ and $v$ ($\overrightarrow{{\tilde{p}-v}}$) all over the mesh. The orientation of the field indicates the position of $\tilde{p}$. The pictures show a detail of the field over a mesh with $40K$ vertices and two mesh sub-samplings with $16K$ and $8K$ vertices: the overall orientation of the field (and therefore the choice of $\tilde{p}$) is robust to different mesh samplings. 
In case of multiple rings, $\tilde{p}$ is selected only on the biggest ring $ring_{N_r}$; for each concentric ring, the starting point is the point $p_i$, which is the closest one to $\tilde{p}$.

Generally the number of elements $p_i \in \mathcal{R}$ varies from one ring to another, because of the increasing radius of the sphere and the irregularity of the tessellation, see Figure \ref{fig:ring_construction}(b). 
To have the same number of elements on every ring, we sample $C$ with $P$ points, where $P$ is a fixed number, called the \emph{spatial resolution}. The results of this sampling is $S$, a set of equidistant samples of $C$, $s_j$ with $j=1, \ldots, P$. 
In details, the equidistant re-sampling is performed as follows:
\begin{itemize}
\item we set the expected distance $\delta r$ between two successive points in $S$ as $\delta r= \dfrac{2\pi R}{P}$;
\item we set $s_0=\tilde{p}$ and extract the points $s_j$ on $C$ such that $$|s_{j-1}-s_j|\approx \delta r, \qquad j=\{1, \ldots, P\}.$$ 
\end{itemize}
The value $h(s_j)$ is linearly approximated from the values the function $h$ assumes on the extrema of the corresponding segment in $C$. 
%
%

\subsubsection{Choice of the ring radii}
With the edgeLBP we are interested to code local variations on the surface, therefore the radius $R$ should be kept small with respect to the overall dimension of the surface.
This implies that the choice of the radius $R$ is crucial for the type (and the size) of the patterns we are going to identify; indeed it must be not too large to avoid to mix global and local surface information and not too small to become insignificant.
In practice, the multiply connected regions appear in case of topological noise, like small handles and mesh self-intersections; in our experiments over thousands of tessellations we never met meaningful admissibility problems. 

We opt for a uniform distribution of the ring radii values. Denoting $R_{max}$ the maximum radius and $N_r$ the number of rings, the value of the ring radii will be $\frac{R_{max}}{N_r}, 2\frac{R_{max}}{N_r}, \ldots, R_{max}$. 

%
%
%
%
\begin{figure*}
\centering
\begin{tabular}{|c|cccc|c|}
\hline
Base models&\multicolumn{4}{c}{Textures}& Textured models\\
\hline
&Class 1&Class 2&Class 3&Class 4&\\
\includegraphics[width=1.4cm]{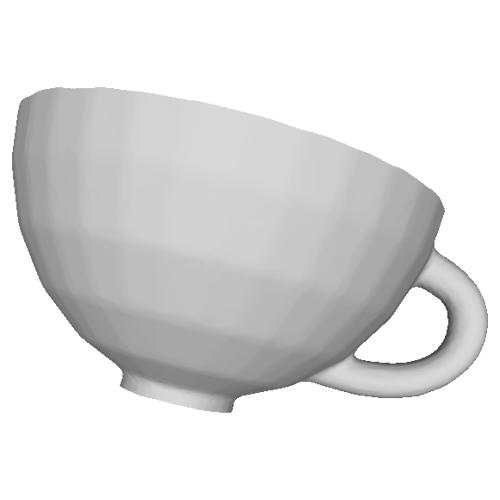}&
\includegraphics[width=1.4cm]{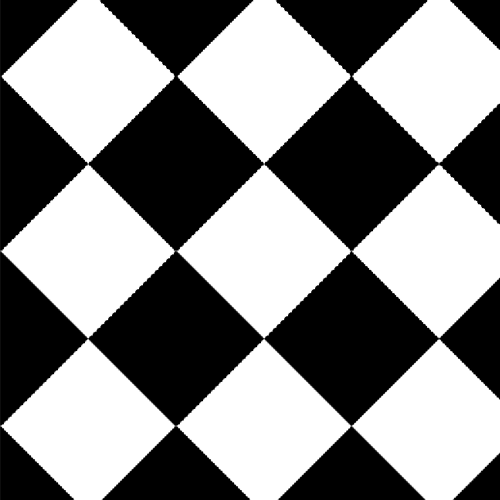}&
\includegraphics[width=1.4cm]{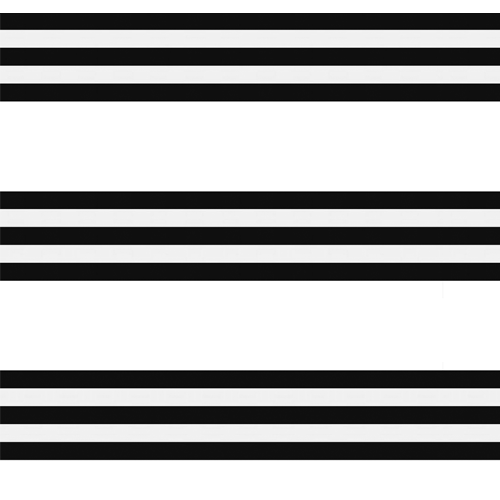}&
\includegraphics[width=1.4cm]{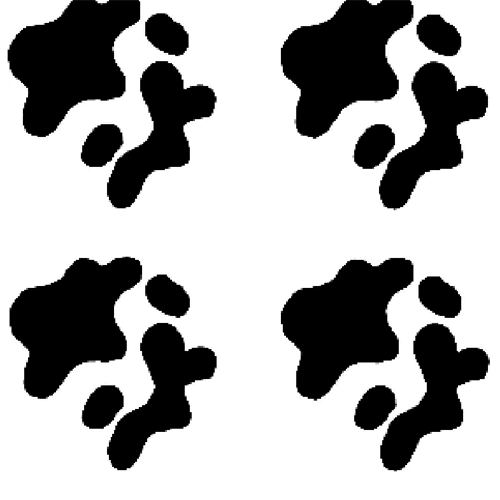}&
\includegraphics[width=1.4cm]{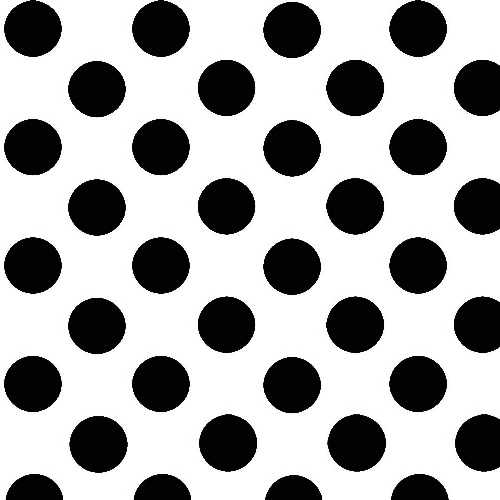}&
\includegraphics[width=1.4cm]{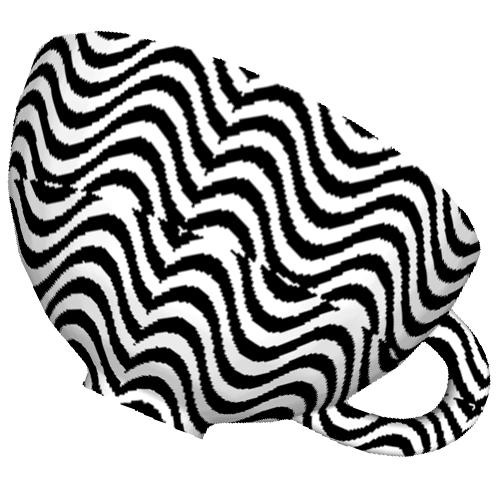}\\
&Class 5&Class 6&Class 7&Class 8&\\
\includegraphics[width=1.4cm]{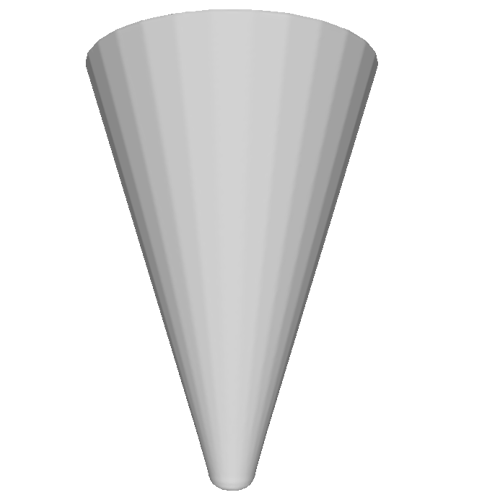}&
\includegraphics[width=1.4cm]{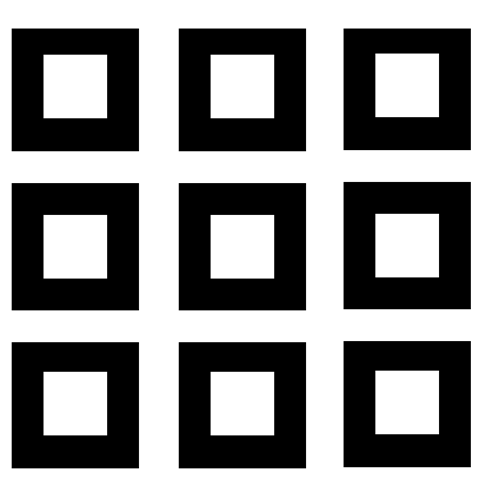}&
\includegraphics[width=1.4cm]{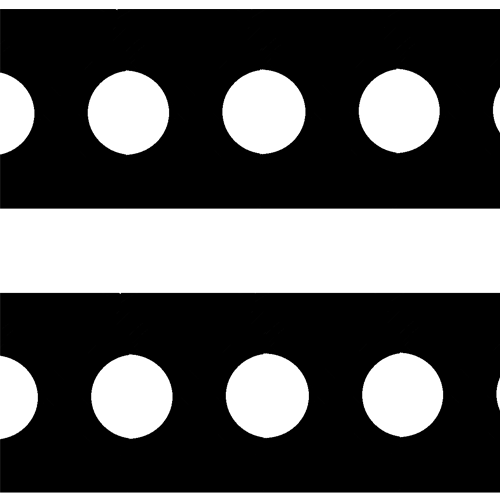}&
\includegraphics[width=1.4cm]{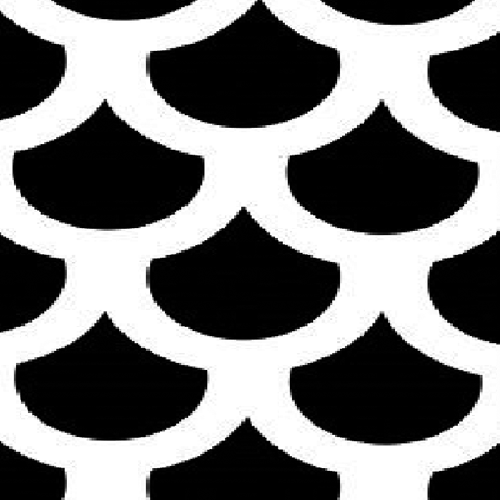}&
\includegraphics[width=1.4cm]{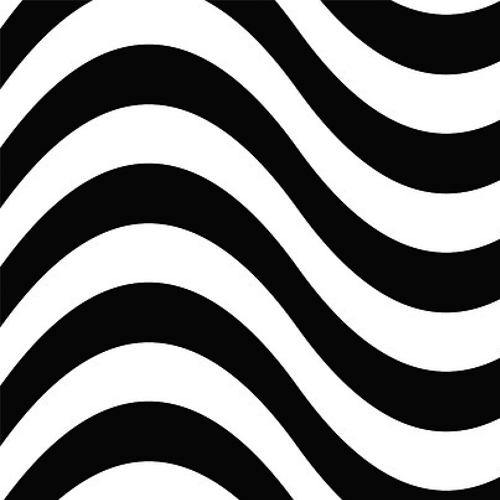}&
\includegraphics[width=1.4cm]{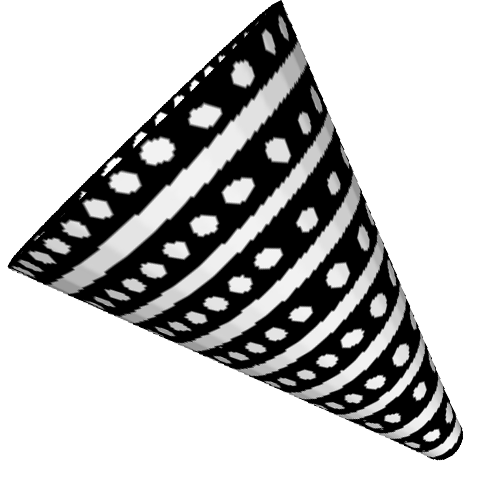}\\
&\multicolumn{2}{c}{Class 9}&
\multicolumn{2}{c|}{Class 10}&\\
&\multicolumn{2}{c}{\includegraphics[width=1.4cm]{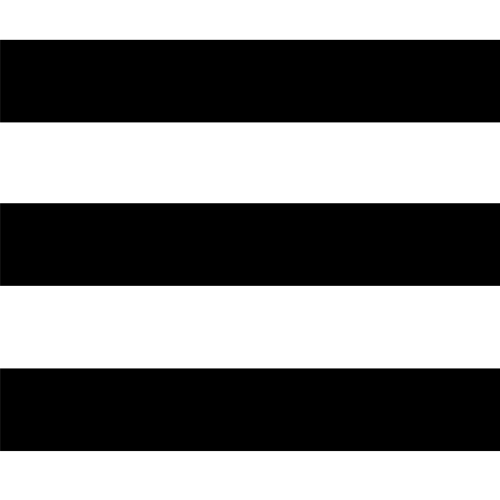}}&
\multicolumn{2}{c|}{\includegraphics[width=1.4cm]{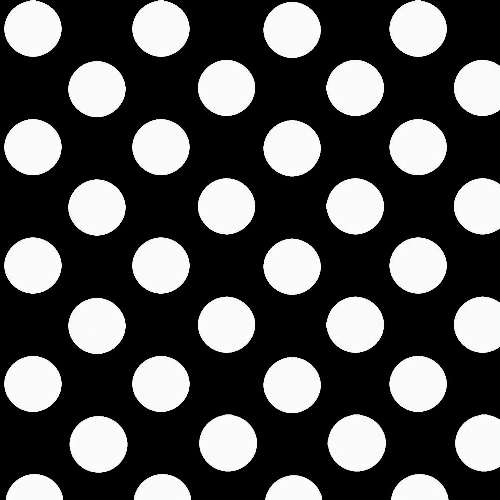}}&\\
\hline
\end{tabular}
\caption{Left: Two of the original models. Center: The ten patterns imprinted on the models of the CPP dataset. Right: Two examples of the textured models of the CCP dataset.}
\label{fig:CCP_origin}
\end{figure*}
\subsection{Similarity assessment}
\label{ss:descriptor}
Once the function $h$ is evaluated over the sample sets of the rings around $v$, the edgeLBP value on $v$ straightforwardly follows from the classic LBP definition, see Section \ref{ss:LBP_imagesc}.

Given the surface tessellation $T$, its \emph{edgeLBP descriptor} is labeled $D_T$. The entry $D_T(n,m)$ is defined as the histogram that counts how many vertices have an edgeLBP value equal to $m$ on the $ring_n$. Since in the experiments we are mostly interested in the distribution of the edgeLBP values, we adopt $\frac{D_T}{n_v}$ as the edgeLBP descriptor, where $n_v$ is the number of the admissible vertices. Through this normalization of $T$ we achieve robustness to the number of vertices of the surface representation.

We define the dissimilarity  between two tessellations $A$ and $B$ as the distance between their corresponding edgeLBP descriptors $D_A$ and $D_B$. 
Since the edgeLBP can be thought as a matrix, any feature vector distance is suitable to evaluate the similarity between two edgeLBP descriptors. We analysed the Euclidean distance between matrices, the \emph{Earth Mover's Distance} as defined in \cite{Rubner:2000} and the \emph{Bhattacharyya distance}. The Bhattacharyya distance $d_{Bha}$ between two distributions $\phi$ and $\psi$ of a scalar random variable $X$ has the following definition:
$$d_{Bha}(\phi,\psi)=\sqrt{1-BC(\phi,\psi)}, \qquad BC(\phi,\psi)=\sum\limits_{x\in X} \sqrt{\phi(x)\psi(x)},$$
where $BC$ is called the \emph{Bhattacharyya coefficient}. Then, for a set of surface tessellations, the dissimilarity values are stored in a \emph{distance matrix} $DM(i,j)=d(D_i,D_j)$, where $d$ is the distance between the descriptors of the tessellation $i$ and $j$. Diagonal values of $Dist(i,i)$ are zero.
\section{Experimental results}
\label{sec:results}
In this Section we introduce the datasets and the evaluation measures adopted to analyse the retrieval performance of the edgeLBP.  We present the edgeLBP performances and discuss its robustness to different tessellations of the same surface.
\subsection{Dataset}
To evaluate the edgeLBP ability of effectively discriminating pattern variations, we used two datasets:
\begin{itemize}
\item the \emph{Cups, Pots and Pans} dataset (or \emph{CPP} for short) is created from triangle meshes in the SHREC'07 Watertight model contest \cite{SHREC07} and the COSEG \cite{COSEG:2012} datasets (see Figure \ref{fig:CCP_origin}(Left)). The original meshes do not have any texture or colorimetric information. From 20 base models and 10 black and white textures representing a pattern (see Figure \ref{fig:CCP_origin}(Center)) we derived 200 models, applying each texture to every model with a semi-automatic algorithm. The proper RGB value was added to the mesh vertices discarding any other colorimetric information (see Figure \ref{fig:CCP_origin}(Right)). At the end of this process, each model is covered by one of the 10 patterns for at least the 30\% of its surfaces while the rest of the surface is only black or only white. The number of vertices of the $200$ models ranges from $95K$ to $107K$. 

\item the \emph{Artifacts} dataset is derived from the laser scans of CH artifacts stored in the STARC repository \cite{Starck07} and selected as test-beds in the Gravitate EU project \cite{GRAVITATE_prog}. The colorimetric information comes as a RGB value associated to each mesh vertex. Differently from the CPP dataset, this second dataset contains full-color information, with a predominance of red, yellow and brown nuances. From these fragments we identified 10 classes of different patterns (see Figure \ref{fig:GRAV_col}); then, for each type of pattern, we tailored 4 representative patches coming from different fragments, for a total of 40 patches. Every patch is made of approximately $40K$ vertices.
\begin{figure}
\centering
\begin{tabular}{ccccc}
Class 1&Class 2&Class 3&Class 4&Class 5\\
\includegraphics[width=2cm]{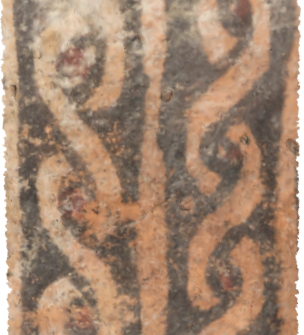}&
\includegraphics[width=2cm]{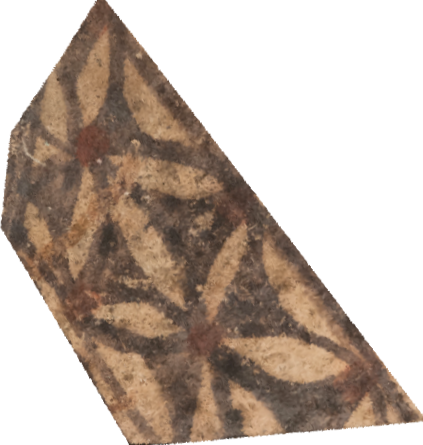}&
\includegraphics[width=2cm]{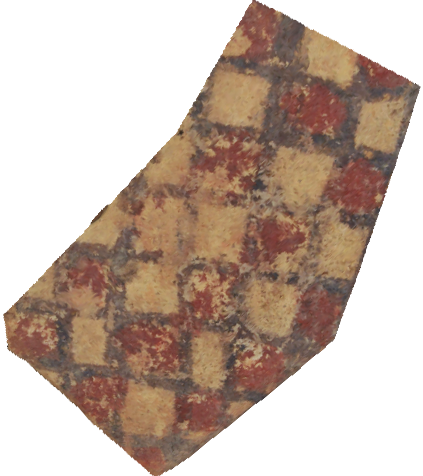}&
\includegraphics[width=2cm]{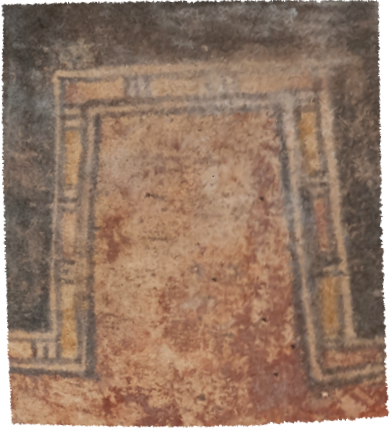}&
\includegraphics[width=2cm]{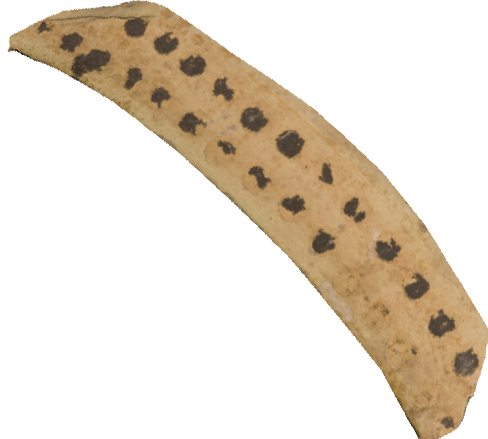}\\
Class 6&Class 7&Class 8&Class 9&Class 10\\
\includegraphics[width=2cm]{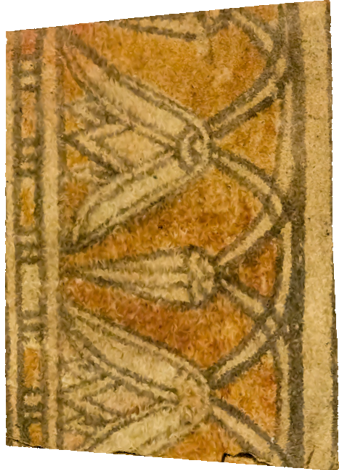}&
\includegraphics[width=2cm]{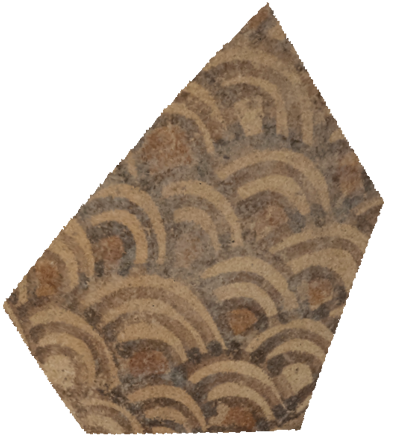}&
\includegraphics[width=2cm]{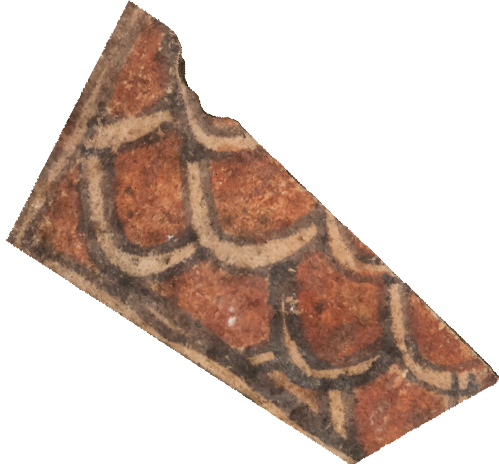}&
\includegraphics[width=2cm]{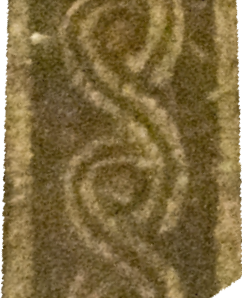}&
\includegraphics[width=2cm]{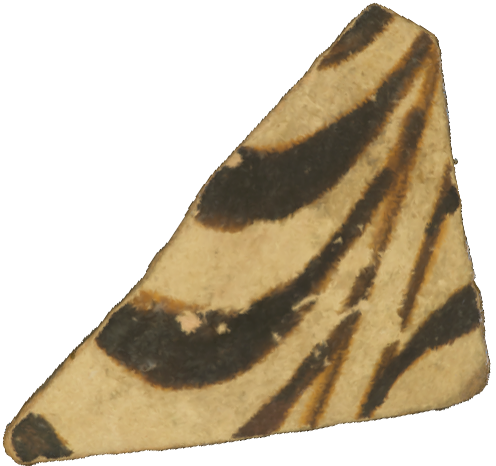}\\
\end{tabular}
\caption{Representatives of the 10 classes considered in the Artifacts dataset.}
\label{fig:GRAV_col}
\end{figure}
\end{itemize}
The edgeLBP algorithm is used to perform colorimetric pattern retrieval on the CCP and Artifact datasets, separately.
\subsection{Evaluation measures}
\label{ss:eval_mes}
The evaluation tests have been performed using a number of classical information retrieval measures, namely the Nearest Neighbor, First Tier, Second Tier, Discounted Cumulative Gain, e-measure, Precision-Recall plot, confusion matrices and tier images.

\paragraph*{Nearest Neighbor, First Tier, Second Tier}
These measures aim at checking the fraction of models in the query's class also appearing within the top $k$ retrievals. 
In detail, for a class with $|C|$ members, $k=1$ for the Nearest Neighbor (NN), $k =|C|-1$ for the first tier (FT), and $k = 2(|C| - 1)$ for the second tier (ST). Note that all these values range from 0 to 1.

\paragraph*{Discounted cumulative gain}
The Discounted Cumulative Gain (DCG) is an enhanced variation of the Cumulative Gain, which is the sum of the graded relevance values of all results in the list of retrieved objects of a given query. The definition of DCG adopted in this paper can be found in \cite{Jarvelin2002}.
\paragraph*{Precision-Recall, mAP and e-measure}
The \emph{Precision} and \emph{Recall} are common measures for retrieval evaluation. 
Recall is the ratio of the number of relevant records retrieved to the total number of relevant records, while precision is the ratio of the number of relevant records retrieved to the size of the return vector \cite{salton_evaluation}. 
Precision and recall always range from 0 to 1. Often, precision and recall are plot as a curve in the reference frame recall vs. precision \cite{Baeza-Yates:1999}: the larger the area below such a curve, the better the performance under examination. As an additional index, we consider the mean Average Precision (mAP), which is the portion of area under a precision-recall curve. Finally, we consider the \emph{e-measure} $e$ \cite{Rijsbergen1979}, which is a quality measure of the first models retrieved for every query.
The e-measure depends on the \emph{Precision} and \emph{Recall} values by the relation: $e=\frac{2}{Precision^{-1}+Recall^{-1}}.$

\paragraph*{Confusion matrices and Tier images}
Each classification performance can be associated with a confusion matrix $CM$, that is, a square matrix whose dimension is equal to the number of classes in the dataset. For the row $i$ in $CM$, the element $CM(i,i)$ gives the number of items which have been correctly classified as elements of the class $i$; similarly, elements $CM(i,j)$, with $j\neq i$, count the items which have been misclassified, resulting as elements of the class $j$ rather than elements of the class $i$. 
Similarly, the tier image $TI$ visualizes the matches of the NN, FT and ST. The value of the element $TI(i,j)$ is: \emph{black} if $j$ is the NN of $i$, \emph{red} if $j$ is among the $(|C|-1)$ top matches (FT) and \emph{blue} if $j$ is among the $2(|C|-1)$ top matches (ST). 
For an ideal classification matrix, $CM$ becomes the diagonal matrix while the $TI$ 
clusters the black/red square pixels on the diagonal.
\subsection{Results}
\label{ss:ris_eval}
In this Section we discuss the retrieval and classification performance of the edgeLBP. For simplicity, we report only the results obtained with the Bhattacharyya distance because in our experiments it performs better than the other distances considered.
%

We performed multiple runs with different settings, changing the number ($N_r$) of rings and the number of samples ($P$) on them, together with different $R$ associated to the $N_r$-th ring (called $R_{max}$). 
The value of $R$ is based on the size of the patterns in it: we randomly picked 3 models of that dataset and choose one or more $R_{max}$ values that were properly scaled for the dataset.
The parameters $N_r$ and $P$ are initially set with what we consider the default settings: $P=15$, $N_r=5$. 
Similarly we consider $h=L-channel$ of the CIELAB color space as the default setting of the function $h$.
Different choices of $h$, $P$ and $N_r$ are discussed for the Artifacts dataset. 

\paragraph*{CPP dataset}
We tested the edgeLBP on this dataset using the default settings and adapting the $R_{max}$ to the size of the wanted pattern ($R_{max}=0.04mm$), in what in this paper is called \emph{Run1}. 
As baseline methods to compare against the edgeLBP descriptor we consider two variations of the color histograms. \emph{Hist1} outputs descriptors based on a 16-bin histogram normalized on his minimal and maximal $L$ values. \emph{Hist2} is similar, but no normalization is applied to the values of $L$.  In addition, we also consider the meshLBP descriptor as implemented in the Matlab toolbox \cite{meshLBP}. 

\begin{figure}

\begin{tabular}{c}
	\begin{tabular}{c}
		\begin{tabular}{|c|c|c|c|c|c|c|}
			\hline
			&NN & FT & ST & e & mAP & nDCG\\
			\hline
			\emph{edgeLBP} & \textbf{0.985} & \textbf{0.801} & \textbf{0.97} & \textbf{0.66} & \textbf{0.859} & \textbf{0.94} \\
			\hline
			\emph{meshLBP} &0.94 & 0.615 & 0.805 & 0.54 & 0.691 & 0.87 \\
			\hline
			Hist1 &0.3 &0.301&0.415&0.27&0.354&0.58\\
			\hline
			Hist2 &0.61&0.522&0.774&0.51&0.57 &0.76\\
			\hline
		\end{tabular}
		\\
		\\
		\includegraphics[width=10cm]{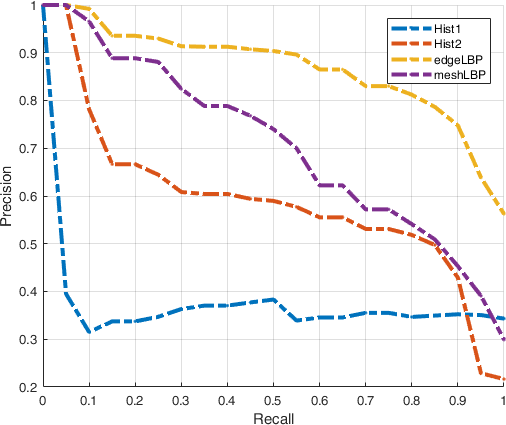}
\end{tabular}
\\
\begin{tabular}{cccc}
	\multicolumn{2}{c}{\Large edgeLBP}&\multicolumn{2}{c}{\Large meshLBP}\\
	Confusion Matrix& Tier Image & Confusion Matrix & Tier Image\\
	\includegraphics[width=3cm]{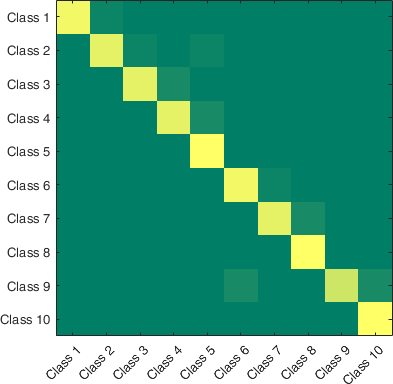}&
	\includegraphics[width=3cm]{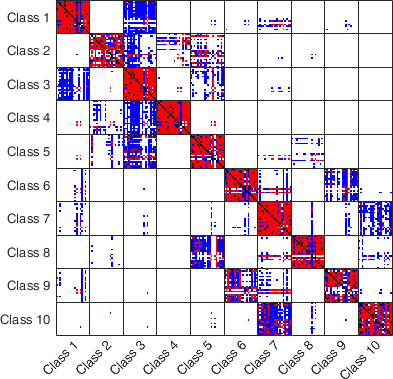}&	
	\includegraphics[width=3cm]{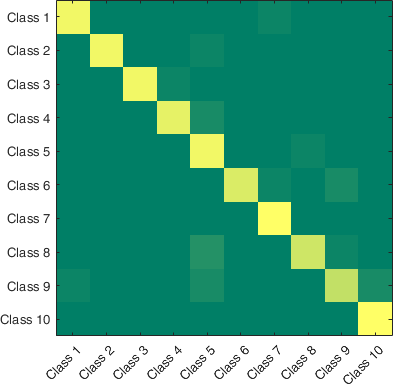}&
	\includegraphics[width=3cm]{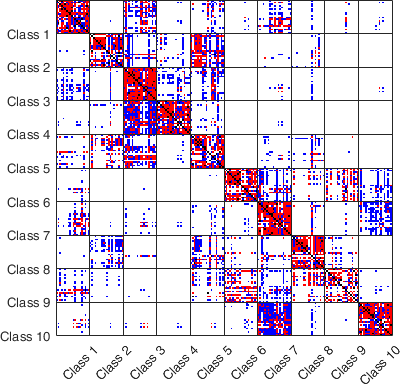}
\end{tabular}
\end{tabular}

\medskip

\caption{Performance evaluation on the CCP dataset.
Top: the NN, FT, ST, e-measure, mAP and nDGC evaluation measures.
Middle: the Precision-Recall curves. Bottom: the confusion matrix and tier image of the edgeLBP and the meshLBP runs.}
\label{fig:ccp_res}
\end{figure}

Figure \ref{fig:ccp_res}(Top) reports the numerical evaluation measures. Figure \ref{fig:ccp_res}(Middle) compares the recall vs precision curves of all the methods. Figure \ref{fig:ccp_res}(Bottom) reports the confusion matrix and the tier image of edgeLBP and the meshLBP runs.
The classification and retrieval results obtained over this dataset are very promising and highlight how the edgeLBP encoding captures the pattern distribution over the surface. The edgeLBP overcome simple histogram-based descriptions that, in practice, measure the percentage of color distribution without any control around vertices and also the meshLBP description that bases the ring definition on mesh elements. The positive edgeLBP perfomance is confirmed in the recent SHREC'18 track for gray color patterns \cite{shrec18}.

%
\subsection*{Artifacts dataset}

\begin{figure*}
\begin{tabular}{cccc}
\hline
\multicolumn{2}{|c|}{\emph{\large $P: 15, N_r:5 R_{max}:0,2$}}&\multicolumn{2}{|c|}{\emph{\large $P: 15, N_r:5 R_{max}:0,5$}}\\
\hline
&&&\\
\includegraphics[width=3cm]{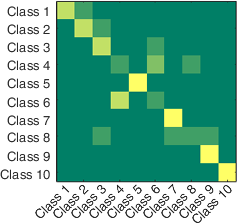}&
\includegraphics[width=3cm]{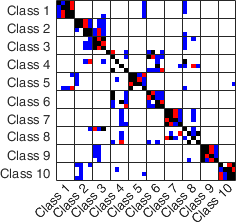}&
\includegraphics[width=3cm]{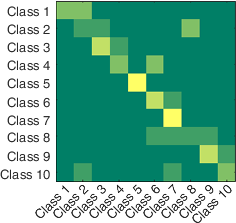}&
\includegraphics[width=3cm]{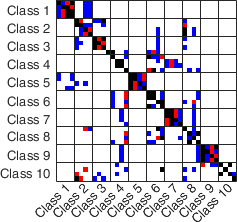}
\end{tabular}
\caption{Confusion matrices and tier images for two of the best runs of the edgeLBP on the Artifacts dataset. In the tier images, the black dots represent the NN element, the red dots correspond to points in FT while blue ones are the ST.}
\label{fig:ar_cmti}
\end{figure*}

This dataset is challenging because of the quality of the original fragments, as their colorimetric patterns are degraded and damaged. Table \ref{tb:GRA_res} reports the NN, FT and ST evaluations for different parameter settings of the edgeLBP. Confusion matrices for the two best radius values are reported in Figure \ref{fig:ar_cmti}, along with the relative Tier Images. The number of models in this dataset is too small to consider meaningful the other evaluation measures.

The edgeLBP achieves good retrieval and classification results for most classes. We observed, as expected, that the correctness of the classification is mainly driven by the size of $R$, rather then $P$ and $N_r$. 
%
%
As a final note, we tested our algorithm using gray scale values as $h$ function: the results obtained with it were pretty much the same as those obtained with $h=L$. We think that this is due to which information both $L$ of CieLAB color space and the gray scale encodes. 
\begin{table}
\centering
\caption{The NN, FT and ST scores for some runs of the edgeLBP on the Artifacts dataset. The $*$ in the fourth row means that in these settings we adopt $h^3$ instead of $h$ (here, $h$ corresponds to the L-channel). $R$ is expressed in $mm$.
}
\begin{tabular}{|c|c|c|c|}
\hline
Parameter Settings					&	NN		&	FT	&	ST\\ 
\hline
$P:15, N_r:5, R_{max}:0,2$				&	0.775	&	0.789	&	1\\
\hline
$P:15, N_r:5, R_{max}:0,3$				&	0.75	&	0.811	&	0.989\\
\hline
$P:15, N_r:5, R_{max}:0,5$				&	0.75	&	0.711	&	0.889\\
\hline
$P:15, N_r:5, R_{max}:0,7*$				&	0.725	&	0.667	&	0.756\\
\hline
$P:12, N_r:7, R_{max}:0,5$				&	0.75	&	0.789	&	0.9\\
\hline
$P:12, N_r:7, R_{max}:0,2$				&	0.775	&	0.856	&	0.978\\
\hline
$P:18, N_r:5, R_{max}:0,7$				&	0.7		&	0.667	&	0.744\\
\hline
\end{tabular}
\label{tb:GRA_res}
\end{table}
%

%
%
%
\subsection{Robustness over different surface tessellations}
\label{ss:robustness_part}
The strength of the edgeLBP is its ring definition, which is robust to different surface tessellations: in this Section we experimentally discuss this robustness. To this aim we re-sample the triangles meshes with a decreasing number of vertices. The triangle mesh re-sampling with $x$ vertices is done with the MeshLAB tool \cite{meshlab} that approximates the original mesh preserving its geometry as much as possible with the given number of vertices (for instance, $x=40K$ vertices).
This process generally modifies the mesh connectivity and the area of the triangles, discards the smallest details and keeps the overall shape, unless the number of vertices drastically diminishes and the new vertices are too few to preserve it.

\begin{figure}
\centering
\begin{tabular}{ccc}
\Large$40K$	&	\Large$24K$	&	\Large$8K$	\\
\includegraphics[width=2.45cm]{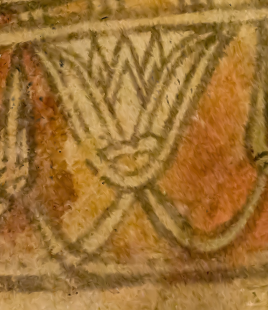}&
\includegraphics[width=2.45cm]{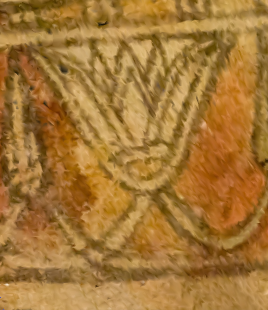}&
\includegraphics[width=2.45cm]{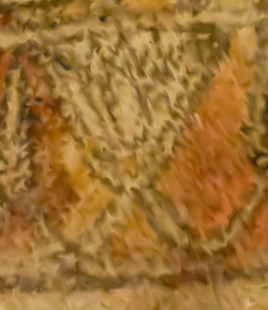}
\end{tabular}
\caption{The degradation of one of the model used to test the robustness of the descriptor of both edgeLBP and meshLBP. The number on each image is the respective vertex resolution.}
\label{fig:comparison_model}
\end{figure}

First, we re-sampled the meshes in the CCP dataset with $40K$ vertices.
On this dataset, we compare the outcome of the edgeLBP with the default settings with the meshLBP, see Table \ref{tb:downgrades}. If compared with the performances on the original CPP dataset in Figure \ref{fig:ccp_res}, the edgeLBP degrades less than the meshLBP, demonstrating of being more robust to mesh degradation and re-sampling.

\begin{table}
\caption{Evaluation measures of the performances on the CPP dataset resampled with 40K vertices.}
\centering
\begin{tabular}{|c|c|c|c|c|c|c|}
\hline
&NN & FT & ST & e & mAP & nDCG\\
\hline
edgeLBP &0.95 	& 0.688 & 0.857 & 0.59 & 0.761 & 0.9\\
\hline
meshLBP &0.77	& 0.517 & 0.703 & 0.47   & 0.58  & 0.79\\
\hline
\end{tabular}
\label{tb:downgrades}
\end{table}

Second, we selected 3 patches from the Artifacts dataset and sub-sampled them with $32K$, $24K$, $16K$ and $8K$ vertices (see Figure \ref{fig:comparison_model}).
These four meshes are compared against the original patch (that has $40K$ vertices).
These four distance values provide an estimate of the error the descriptors do when working with the simplified meshes. 

\begin{figure}
\centering
\begin{tabular}{c}
\emph{Run1}\\
\includegraphics[scale=.9]{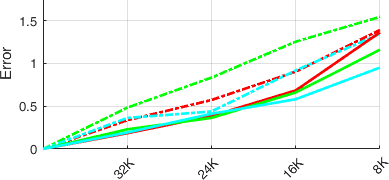}\\
\emph{Run2}\\
\includegraphics[scale=.9]{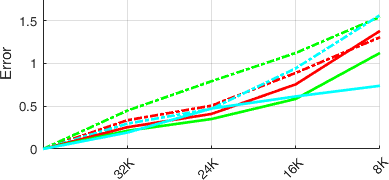}\\
\includegraphics[scale=.9]{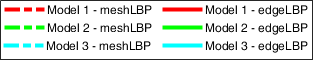}\\
\end{tabular}

\caption{The plots represent the distance of the four simplified meshes from the original ones, with respect to the meshLBP and the edgeLBP descriptors. The labels in the horizontal axis highlight to the number of vertices of the mesh.}
\label{fig:mesh_edge_results}
\end{figure}

We performed two runs for both the edgeLBP and meshLBP:
\begin{itemize}
\item \emph{Run1: $P=12$, $N_r=7$}.  These settings are the setting used by the meshLBP as default. Both meshLBP and edgeLBP are run with these settings. For the edgeLBP we set $R_{max}=0.5mm$. 
\item \emph{Run2: $P=15$, $N_r=5$}. These settings are those that we consider default for the edgeLBP. Both the algorithms are run with these settings. As in $run1$, we set $R_{max}=0.5mm$.
\end{itemize} 
Figure \ref{fig:mesh_edge_results} represents the distance between the original model and its four approximations with respect to both edgeLBP and meshLBP, for all the three original meshes. 
Since the scale of the distances adopted by the meshLBP and edgeLBP is different, we normalize them with respect the range of the distance values among these patches.
%
From Figure \ref{fig:mesh_edge_results}, we can see that in both runs the edgeLBP produces more stable descriptors, 
as the errors are lower than those of the meshLBP (except in one case, the model 1 in \emph{Run2}). In our opinion the nature of the ring definition of the two methods is crucial being both methods based on the LBP concept. Indeed, the meshLBP creates rings of different size when the vertex density decreases becoming quite sparse when the number of vertices of the mesh is significantly reduced. 
This is not the case of the edgeLBP, as the radius of each ring is always the same ($R$), for each mesh.

%
\section{Discussions and conclusive remarks}
\label{sec:conclusions}
We defined an extension of the LBP on surfaces, whose strength is the robustness to the surface tessellation. In this paper we used this technique to successfully retrieve and classify colorimetric patterns on mesh surfaces. The edgeLBP also performed the best to the SHREC'18 track on retrieval of colorimetric patterns \cite{shrec18}. Besides synthetic datasets, we tested our algorithm on samples coming from a challenging dataset made of corrupted and degraded artifacts of the EU GRAVITATE project test beds \cite{GRAVITATE_prog}, achieving
promising results.
%
Further extensions are planned and possible. 
For instance, it is possible to adopt this approach for the description of geometric patterns, encoding the geometric variations with scalar properties of the mesh, like mean curvature or shape index.
Moreover, we think that for full color patterns better results could be achieved using all the colorimetric information, for instance the L, a, and b channels of the CIELab space. In this direction, we are currently working on the extension of the edgeLBP to multidimensional properties. 

Finally, we think it is worth investigating the automatic recognition and localization of multiple patterns on surfaces. Current experiments are performed on surfaces fully characterized by a single pattern at a time and the similarity distance is defined on the global fragment skin.
Next plans include the combination of the shape description step with segmentation techniques and the aggregation of parts made of vertices with similar local descriptions.
\section*{Acknowledgments}
The work is developed within the research program of the ``H2020'' European project ``GRAVITATE'', contract n. 665155, (2015-2018).

\bibliographystyle{abbrv}
\bibliography{bibliography}
\newpage

\end{document}